\documentclass[12pt]{JHEP3}
\usepackage{epsfig}
\usepackage{amsmath}
\def\beq{\begin{eqnarray}}
\def\eeq{\end{eqnarray}}
\title{
Glueball Decay in Holographic QCD}
\author{
Koji Hashimoto$^{\dagger a}$, Chung-I Tan$^{+ b}$ and 
Seiji Terashima$^{* c}$\\
${}^\dagger$ {\it Institute of Physics, the University of Tokyo, Komaba,
Tokyo 153-8902, Japan}\\
${}^+$ {\it Department of Physics, Brown University, Providence, RI
02912, USA}\\
${}^*$ {\it Yukawa Institute for Theoretical Physics, \\
\hspace{40mm}Kyoto University, Kyoto 606-8502, Japan}\\
$^a$ E-mail: \email{koji@hep1.c.u-tokyo.ac.jp}\\
$^b$ E-mail: \email{tan@het.brown.edu}\\
$^c$ E-mail: \email{terasima@yukawa.kyoto-u.ac.jp}\\
}

\abstract{
Using  holographic QCD based on D4-branes and D8-anti-D8-branes,
we have computed couplings of glueballs to light mesons. 
We describe glueball decay by explicitly calculating 
its decay widths and branching ratios.
Interestingly, while  glueballs remain less well  understood 
both theoretically and experimentally,
our  results are found to be
consistent with the experimental data for  the scalar 
glueball candidate $f_0(1500)$.
More generally, holographic QCD predicts that decay of any glueball to
4$\pi_0$ is 
surpressed, and that mixing of the lightest glueball with $q\bar{q}$ 
mesons is small.}

\preprint{
{\normalsize Brown-HET-1489}\\
{\normalsize UT-Komaba/07-14}\\
{\normalsize YITP-07-54}
}

\begin{document}

\section{Introduction}

Glueballs,  as excitations of 
gauge-invariant composite operators in Yang-Mills theories, 
remain  illusive. 
Although the existence of the glueballs (of various types, such
as scalar glueballs and tensor glueballs) is expected, their experimental
identification in the hadron spectra remains 
difficult.\footnote{For details, we refer readers to the section
of  ``Non-${\rm q} \bar{\rm q}$ candidates'' in Meson Particle Listings
in \cite{Yao:2006px}, and to \cite{Amsler:2004ps} 
for recent discussions.} This difficulty  is
largely due to the inability to compute reliably couplings of glueballs
to ordinary mesons in strongly coupled QCD. Lattice QCD predicts for the
mass of the lightest scalar glueball to be around 1600-1700 MeV 
\cite{Morningstar:1999rf,Lee:1999kv}, but
it doesn't yet provide information  on the glueball couplings and decay
products/widths, which are indispensable for their identification. 
The Large Hadron Collider (LHC) will likely yield a huge amount of
hadronic data, which can lead to  progress in
revealing the mystery of glueballs.

In this paper, we explicitly compute the couplings between light 
glueballs and light $q\bar{q}$
mesons, by using holographic QCD. AdS/CFT (gauge/gravity) 
correspondence (duality)
\cite{Maldacena:1997re,Gubser:1998bc} is one of the most important
developments in string theory, and,   holographic QCD refers to the
application   of AdS/CFT to QCD studies. The basic claim of the AdS/CFT
correspondence is that  
correlation functions of gauge-invariant composite operators in large
$N_c$ gauge theories at strong 't Hooft coupling correspond to
classical gravitational computations in higher dimensional gravity
theories in curved backgrounds. The correspondence has been applied to
(i) 
computation of glueball spectrum  in large $N_c$ 
pure Yang-Mills theory and
to (ii) $q\bar{q}$ 
meson spectra/dynamics in large $N_c$ QCD, which we review briefly below. 
These efforts  have been quite successful in reproducing lattice and 
experimental data of hadrons,
even though the real QCD is recovered in the ``CFT'' side 
only when one incorporates various corrections in the 
large $N_c$ and large 't Hooft coupling expansion.
Here we combine these two efforts, (i) and (ii),
in order to calculate 
couplings between the glueballs and the $q\bar{q}$ mesons, in the large
$N_c$ QCD. 

The key merit of using  holographic QCD is the fact that one not only
can calculate the hadron spectra, but also can compute explicitly their
{\it couplings}.  It provides a more powerful  method for constraining
these couplings  than the chiral perturbation technique. In particular,
since glueballs are expected to be heavier than 1 GeV, derivative
expansion in chiral perturbation becomes unreliable.  Furthermore,
current  lattice calculations  are not well suited for computing
dynamical quantities such as 
decays and couplings. Our paper represents a first principle calculation
for  glueball decays, though in the approximation where the holographic
duality is valid. 

Let us briefly review  here the holographic study of the two sectors (i)
and (ii).  
Glueball  studies began at the early stage of  AdS/CFT correspondence,
since they should exist in pure 
Yang-Mills theories  whose supersymmetric version was the basic building
block of 
the correspondence. Witten was the first to  suggest a reliable way in
breaking the 
supersymmetries thus allowing one to   treat  the  bosonic Yang-Mills
theory \cite{Witten:1998zw}.  
The gravity dual is the near horizon limit of 
a classical solution of 10 dimensional type IIA supergravity
representing $N_c$ D4-branes wrapping an $S^1$
with anti-periodic boundary condition for fermions. 
After various developments along this direction 
\cite{Csaki:1998qr,Gross:1998gk,Constable:1999gb,Brower:1999nj}, 
a complete  spectrum of scalar/tensor glueballs in four-dimensional
Yang-Mills 
theory was given in \cite{Brower:2000rp}  where they appeared as
graviton/dilaton/tensor fluctuations 
in the Witten's gravity background. 
The calculated  glueball spectrum is consistent with the  lattice
computations, 
and in this paper, we compute glueball decays, based on this spectrum
\cite{Brower:2000rp}. 
The lightest glueball is a scalar state with quantum numbers 
$J^{PC}=0^{++}$. Since this state  should be the easiest one to identify
in hadronic 
data, we concentrate here on this lightest scalar glueball for explicit
computations. The gravity dual of this lightest scalar glueball
corresponds to a specific  
combination of  metric fluctuations.

The quark sector, (ii),
is obtained in AdS/CFT correspondence by the  introduction of
flavor D-branes \cite{Karch:2002sh}
intersecting with color $N_c$ D4-branes. 
Various D-brane configurations (and also phenomenological holographic 
models) describing flavor/chiral physics have been proposed (see 
\cite{Kruczenski:2003be,Son:2003et,Kruczenski:2003uq,
Sakai:2004cn,Sakai:2005yt} for a partial list).  Among these, 
we shall use the 
Sakai-Sugimoto model \cite{Sakai:2004cn,Sakai:2005yt},
which has been  quite successful in reproducing various
facets of low energy QCD dynamics while maintaining  its
string-theoretical  
origin. The Sakai-Sugimoto model uses $N_f$
D8-branes and $N_f$ anti-D8-branes as the flavor D-branes, and
their intersection with color $N_c$ D4-branes gives rise to 
string excitations corresponding to the quarks. Among various ways of
introducing flavor D-branes, the Sakai-Sugimoto model beautifully
realizes spontaneous chiral symmetry breaking and chiral
dynamics in QCD.\footnote{
Although the quarks of the model 
are massless (and so the pions are massless), 
it does not present a problem for our purpose. For recent discussion on
obtaining  massive pions, see \cite{Hashimoto:2007fa}.}
The resulting low energy theory for the quark sector is dual to 
the probe D8-brane worldvolume
theory (higher dimensional Yang-Mills theory) 
in the Witten's supergravity background. 
In particular, the $q\bar{q}$ mesons 
are described as Kaluza-Klein (KK) 
decomposed massless fields on the probe
D8-brane. 

We would like to compute  the couplings between the glueballs and the
$q\bar{q}$ mesons in this setting. In the dual description through the
AdS/CFT,  
they correspond to  the supergravity fluctuations and the
Yang-Mills fluctuations on the D8-branes, respectively. These two
sectors are coupled in the  combined system of  supergravity plus
D8-branes. We substitute the fluctuations (wave functions) of the
supergravity fields 
(corresponding to the glueball) and the D8-brane massless fields
(mesons) 
into the D8-brane action and integrate  over the extra dimensions, 
to obtain the desired couplings. 
Combining sectors (i) and (ii) is important not only due to
its phenomenological impact but also because 
 this represents  the first computation   in holographic QCD of 
the couplings between the
supergravity fluctuations   and the fields on the probe D-branes.

Once the couplings are obtained, we can compute the decay widths for
various decay channels of a glueball, and study its possible mixings. 
Because the whole $q\bar{q}$ meson sector is combined 
into the D8-brane action, which is a
higher-dimensional Yang-Mills lagrangian, several interesting
mesonic features
follow.\footnote{One is the reproduction of the vector-meson dominance 
as shown in \cite{Sakai:2005yt}.} 
For example, at the leading order (in the expansion of the large
't Hooft coupling),  glueball 
decay to 4$\pi_0$ is prohibited. There is no
direct 4$\pi_0$ coupling to the glueball, and, furthermore,  glueball -
$\rho$ meson coupling also does not allow  the 4$\pi_0$ decay mode. 
As for the mixing,
we can show that the lightest glueball has no mixing with $q\bar{q}$
mesons at the leading order of our expansion.
These are our main  predictions based on the holographic QCD.

The organization of this paper is as follows.  In section \ref{sec:2},
we explicitly compute the glueball couplings in 
the holographic QCD, and obtain the interaction lagrangian. 
We study generic features of the glueball decay following from the
holographic QCD. 
In section
\ref{sec:3}, we compute the decay widths based on these interactions. 
We list possible decay products, and obtain widths for various allowed
decay 
modes.  
We next  compare these with the experimental data. It has been argued
that  
$f_0(1500)$ is the most plausible candidate for the lightest scalar
glueball \cite{Yao:2006px}, 
and we find that our results are consistent with the
hadronic data for $f_0(1500)$. We reproduce the narrow
width of the $f_0(1500)$, and also the decay products/branching ratio,
qualitatively.
In section \ref{sec:4}, we provide  a summary, discussions, and a list
of directions for 
future studies.

\section{Glueball Interaction}
\label{sec:2}

Holographic QCD, in particular the Sakai-Sugimoto model, has provided  a
novel 
unified view of the mesons in a multi-flavored QCD. 
All the mesons appear just as KK decomposed massless 
fields living in higher dimensions. As a consequence, there
exist many interesting relations among couplings between the
mesons. The Skyrm term is one example. Here our concern
is with the glueballs,  which live in a different sector in the dual
side, {\it 
i.e.} in the supergravity fluctuations, not on the flavor D-branes.
But the holographic features found in the meson sector 
are inherited also for the glueball-meson
couplings. This is because these couplings are also controlled by the
flavor D-brane action, and shares the same flavor structure (commutator
structure of 
the non-Abelian massless fields on the D8-branes). In this section,
after reviewing 
the dual descriptions of the glueballs and $q\bar{q}$ mesons, we
describe the generic features for  glueball decays dictated by the
holography. Finally, using the holographic QCD, 
we explicitly derive the couplings between the lightest scalar
glueball and light $q\bar{q}$ mesons.

\subsection{Brief review of holographic QCD: glueballs and mesons}

\subsubsection{Glueball sector}

Glueballs are gauge-invariant composite states in Yang-Mills theory,
and their duals are fluctuations in near horizon geometry of
black-brane solutions. There are numerous ways to break supersymmetries
by deforming the $AdS_5\times S^5$ solution with 
which the original AdS/CFT
correspondence was derived. Among them, 
the Witten's background \cite{Witten:1998zw} is ``reliable'' in
the sense that it  
knows how the supersymmetries are broken in the Yang-Mills side:
anti-periodic boundary condition for the fermions on the D4-brane
worldvolume. 

Let us review briefly  the description of the gravity 
dual for the lightest glueball in the four-dimensional QCD. 
It corresponds to  supergravity fluctuations in the Witten's
classical background in 10 dimensions. The type IIA supergravity
solutions can be written conveniently in the notation of the 
11 dimensional supergravity, and in that notation the Witten's solution
is a doubly Wick-rotated $AdS_7$ blackhole, 
\begin{eqnarray}
 ds^2 = \frac{r^2}{L^2}\left(
f(r)d\tau^2 + \eta_{\mu\nu} dx^\mu dx^\nu
\right)
+ \frac{L^2}{r^2} (f(r))^{-1} dr^2 + \frac14 L^2 d\Omega_4^2
\label{mmetric}
\end{eqnarray} 
where $f(r)\equiv 1-R^6/r^6$. $L$ and $R$ are the parameters of the
solution. $\mu,\nu$ run from 0 to 4, and the $x^4$ direction is the 11th
dimension (the M-theory circle). The $\tau$ direction is compactified to
a circle, and its radius is fixed as $L^2/(3R)$ so that the
background is non-singular, and  the manifold is smooth,
around the ``end'' of the spacetime  solution (at $r=R$).

The $S^4$ part is not necessary in the
following discussion, so we integrate that part (and also the flux) 
to obtain the M-theory supergravity action reduced to 7 dimensions: 
\begin{eqnarray}
 S = \frac{-1}{2\kappa_{11}^2} \frac{L^4}{16}V_4
\int\! d^7x \sqrt{-\det G_{MN}} \left(R(G_{MN}) + \frac{30}{L^2}\right).
\label{s11-7}
\end{eqnarray}
Here $V_4\equiv 8\pi^2/3$ is the volume of a unit $S^4$, 
and we have  followed the notation of \cite{Constable:1999gb}.\footnote{ 
Further integration of the $\tau$ and the $x^4$ (M-theory circle) gives
a 5 dimensional AdS gravity action, 
\begin{eqnarray}
 S = \frac{-L^4 V_4}{48 (2\pi)^6 l_s^8 g_s^2 R}
\int\! dr d^4x\; r^2 \sqrt{f}\sqrt{-\det G_{MN}}
\left(R(G_{MN}) + {\frac{30}{L^2}}\right)
\label{s11}
\end{eqnarray}
Here $M,N$  run through $(0,1,2,3,r)$, and we have used the M-theory
$\leftrightarrow$ type IIA relations 
$R_{11}= g_s l_s$ and $2\kappa_{11}^2 = (2\pi)^8 l_s^9 g_s^3$.}

In \cite{Brower:2000rp}, 
a complete bosonic  spectrum was given, 
and the lightest state has  quantum numbers 
$J^{PC}=0^{++}$ in terms of the $x^0,\cdots,x^3$ spacetime.
The metric fluctuations for this lightest state 
in the action (\ref{s11-7})
were explicitly obtained in 
\cite{Constable:1999gb}\footnote{
For the analogue  state  in three-dimensional QCD, see 
\cite{Brower:1999nj}.}\footnote{
In \cite{Constable:1999gb} there is a typo in equations
(38) and (40) 
(the sign of the functions $c$ and $b$
is opposite). 
} as
\begin{eqnarray}
 &&
h_{rr} = -\frac{L^2}{r^2}f^{-1} \frac{3R^6}{5r^6-2R^6}
H(r) G(x) , \quad 
h_{44} = \frac{r^2}{L^2}
\frac14
 H(r)G(x), \nonumber \\
&&
h_{\mu\nu} = \frac{r^2}{L^2}
\left(
\eta_{\mu\nu} \frac14 H(r)
- 
\left(
\frac14 + \frac{3R^6}{5r^6-2R^6} H(r)
\right)
\frac{\partial_\mu \partial_\nu}{M^2}
\right)
G(x),
\label{glfl}
\\
&&
h_{r\mu} = \frac{360 r^7 R^6}{M^2 L^2 (10 r^6 - 4 R^6)^2} 
H(r)\partial_\mu G(x), \quad h_{\tau\tau} = -\frac{r^2}{L^2}fH(r)G(x).
\nonumber
\end{eqnarray}
Here $G(x^0,\cdots,x^3)$ is the glueball field in the real
1+3 dimensional spacetime, and $M$ is the mass of the
glueball.\footnote{The excitation tower for these graviton-dilaton 
fluctuations is denoted as $S_4$ in \cite{Brower:2000rp}.}

The mass squared for this glueball state was found in
\cite{Brower:2000rp} 
to be  $M^2 = 7.308 R^2/L^4$,
by solving the eigen-equation following from the equation of motion of
the 7-dimensional AdS supergravity (\ref{s11-7}) given by  the
mass-shell condition of the glueball field $(\square-M^2)G=0$, 
\begin{eqnarray}
- \frac{d}{dr}(r^7-rR^6) \frac{d}{dr}H(r)
-\left(
L^4 M^2 r^3
+ \frac{432 r^5R^{12}}{(5r^6-2R^6)^2}
\right)H(r)=0.
\end{eqnarray}
For our
later purpose, it is useful to change the coordinate to a dimensionless
$Z$ defined by 
\begin{eqnarray}
 r/R = K^{1/6}, \quad K \equiv 1+Z^2.
\label{relz}
\end{eqnarray}
$Z=0$ corresponds to the bottom of the background $r=R$,  and
the branch $Z (\geq 0)$ is smoothly connected to the  branch 
$Z (\leq 0)$ \cite{Sakai:2004cn}. 
In this $Z$ coordinate, the eigen-equation becomes
\begin{eqnarray}
- \frac{3}{Z} \frac{d}{dZ}\!
\left(
\!3Z(1\!+\!Z^2)\frac{d}{dZ}H(Z)\!
\right)
- \left(
\!\frac{L^4M^2}{R^2} (1\!+\!Z^2)^{-\frac13}
 + \frac{432}{(5Z^2\! +\! 3)^2\!}
\right)\!H(Z)=0.\quad
\label{Hd}
\end{eqnarray}
The appropriate boundary condition for solving this is
\begin{eqnarray}
 \frac{d}{dZ}H \biggm|_{Z=0} = 0,
\quad
H(Z=0) \neq 0,
\quad 
H(Z=\infty) = 0.
\end{eqnarray}
Other fluctuations (such as the state corresponding to $2^{++}$
glueball) can be constructed in the same manner 
\cite{Constable:1999gb,Brower:2000rp}. 

\setcounter{footnote}{0}
\subsubsection{$q\bar{q}$ meson sector}

The dual of the quark sector is the probe flavor D-branes intersecting
with the color D-branes. To make sure that we are not simply
constructing a 
phenomenological model but ``deriving''
the hadronic interactions from the first principle, we need to
follow so-called top-down approach from string theory. The
Sakai-Sugimoto model \cite{Sakai:2004cn} is 
the best known top-down construction for multi-flavor quarks.

The $q\bar{q}$ mesons are
described in the Sakai-Sugimoto model 
by the flavor D8-brane action in the Witten's background
(written with type IIA string metric), 
\begin{eqnarray}
 S_{\rm D8} 
= -(2\pi\alpha')^2{\cal T}_{\rm D8}{\rm Tr}\!\int\! d^9x \;
e^{-\Phi}\! \sqrt{-\det \tilde{g}}\; \frac14 \;
\tilde{g}^{PR}\tilde{g}^{QS}
F_{PQ}F_{RS} + S_{\rm Chern-Simons}.
\label{d8ac}
\end{eqnarray}
Here $\tilde{g}_{PQ}$ is the metric induced on the D8-brane worldvolume
spanning the directions $0,1,2,3,r,S^4$, and we have already expanded
the Dirac-Born-Infeld (DBI)action to the second order in the Yang-Mills
field strength.\footnote{There is a tadpole for closed string modes, 
but in this paper we neglect the back-reaction to the metric due to the
presence of the D8-brane because it will be small in the large $N_c$
limit.}   
The normalization of the generators of the gauge group is chosen
as ${\rm Tr} T_a T_b=\delta_{ab}$. 
The parameters in the 
background metric (\ref{mmetric}) are related to the notation of the 
Sakai-Sugimoto model by $L = 2R_{\rm SS}$, 
$R = 2 \sqrt{R_{\rm SS}U_{\rm KK}}$ and (\ref{relz}), 
where $R_{\rm SS}$ denotes ``$R$'' in the original 
papers of Sakai and Sugimoto
\cite{Sakai:2004cn,Sakai:2005yt}. The typical mass scale
appearing in all the computations in \cite{Sakai:2004cn,Sakai:2005yt}
is $M_{\rm KK}\equiv (3/2) U_{\rm KK}^{1/2} R_{\rm SS}^{-3/2}$.

Again, we integrate out the irrelevant $S^4$ part, leading to 
\begin{eqnarray}
 S_{\rm D8} = \frac{-{\cal T}_{\rm D8}(2\pi\alpha')^2 V_4}{4 g_s}
{\rm Tr}\int\! d^4x dz 
\left[
3 R_{\rm SS}^{3/2} U_{\rm KK}^{5/2} K\;
\eta^{\mu\nu}F_{\mu z}F_{\nu z} 
\right. \hspace{20mm}
\nonumber \\
\hspace{50mm}\left. + 
\frac23 K^{-1/3} R_{\rm SS}^{9/2}U_{\rm KK}^{-1/2} \eta^{\mu\rho}
\eta^{\nu\sigma}F_{\mu\nu} F_{\rho\sigma}
\right],
\label{d8red}
\end{eqnarray}
where $z\equiv U_{\rm KK}Z$ and $\mu,\nu=0,1,2,3$. 
(The Chern-Simons term in (\ref{d8ac}) will be irrelevant to our
discussion of the glueball decay;  see section \ref{sec:3}.)
The KK decomposition
along $z$ (equivalently $r$, $Z$) in \cite{Sakai:2004cn,Sakai:2005yt}
is 
\begin{eqnarray}
 A_z = \phi_0(z) \pi(x^\nu), \quad
A_\mu = \psi_1(z) \rho_\mu(x^\nu)
\label{pirhodef}
\end{eqnarray}
where we suppress all the other higher components, since we are
interested in 
the decay of the glueball to light mesons. The eigenfunction
$\psi_1$ should satisfy the eigen-equation which  follows from the
action (\ref{d8red}), 
\begin{eqnarray}
 -K^{1/3} \partial_Z (K \partial_Z \psi_1) = \lambda_1 \psi_1
\end{eqnarray}
where $\lambda_1 = 0.669$ 
for normalizable $\psi_1$, leading to  the mass squared
for the $\rho$ meson, $m_\rho^2 = \lambda_1 M_{KK}^2$.
The pion is massless, with its eigenfunction $\phi_0(z)\propto 1/K$.

The trace in (\ref{d8red})
is for the matrix-valued mesons. For $N_f$ flavors, the pions
and the $\rho$ mesons are $N_f\times N_f$ matrices. The overall trace
part of the pion should yield  a mass from the chiral anomaly (see
\cite{Sakai:2004cn,Sakai:2005yt} for the description of the 
supergravity counterpart) but it is not included here. 
In this paper we take $N_f=2$,
hence the overall trace part of the pion is $\eta$ (or $\eta'$) meson, 
while that of the $\rho$ meson is $\omega$. Except when adjoint indices
are explicitly written, we include $\eta$ ($\eta'$) and $\omega$ in
the matrix notation of the fields $\pi$ and $\rho$. 

\subsection{Generic features of holographic glueball decay}
\label{sec:gene}

Our strategy to compute the interaction between the glueballs and the
$q\bar{q}$ mesons is very simple. Since we know how all these hadrons
are described in the dual side (as in (\ref{glfl}) and
(\ref{pirhodef})), we substitute them into the D8-brane action
(\ref{d8ac})  and integrate it over the extra dimensions. In the
original Sakai-Sugimoto model, the induced metric $\tilde{g}$ in
(\ref{d8ac}) was just the background metric, but now the glueball
appears as a fluctuation in the induced metric and the dilaton in the
D8-brane action.

Since the appearance of the glueball doesn't break the non-Abelian
structure of the
D8-brane action, we can expect that some generic features of the
glueball coupling may be read in the D8-brane action. As an obvious
check,
the glueballs should be flavor-blind; This can be seen as the fact that 
the supergravity fields are gauge invariant with respect to the
gauge transformation on the D8-brane. As a consequence, couplings of the
glueballs to $q\bar{q}$ mesons are universal against flavors.

Gauge invariance in  higher dimensions also constrains the meson
interactions. For example, as Sakai and Sugimoto have shown, the 
Skyrm term in the pion self-interactions is 
encoded in the structure of the higher dimensional Yang-Mills
lagrangian. In our case, even though  glueballs (=gravity and dilaton
fluctuations) are now included, this flavor structure is almost
unaltered. We note the following  interesting features: 
\begin{itemize}
 \item[(a)] There are no glueball interactions involving  more than two
	    pions.  
\item[(b)] For glueball coupling to $\rho$ and pion, the $\rho$ meson
	    couples to the pion as if the pion were charged under the
	    $\rho$  meson gauge field. 
\item[(c)] Direct coupling of a glueball, $G$,  with more than five
	    mesons are suppressed by large 't Hooft coupling.
\end{itemize}
(a) is easily seen by noting the fact that $\pi$ appears
in $A_z$ but there are no  $(A_z)^n$ terms, with $n>2$ 
in the D8-brane action (\ref{d8ac}). 
(b) follows from the fact that $F_{\mu z}\sim [A_\mu,A_z]$ can be 
decomposed as $[\rho_\mu, \partial_\mu \pi]$.
These two features are precisely what were observed in the
Sakai-Sugimoto model for the pure $q\bar{q}$ meson sector, and now
inherited to the glueball couplings.
Finally, (c) is due to the fact that the higher dimensional 
Yang-Mills action (\ref{d8ac})
does not have $A^5$ term. For (a) and (c), DBI corrections give $F^4$
terms but they are suppressed by $\alpha'$ or equivalently the large
't Hooft coupling. 

To be more explicit, we can list the couplings which appear in
(\ref{d8ac}): 
\begin{eqnarray}
G {\rm Tr}(\pi^2), \;
G{\rm Tr}(\pi,[\pi,\rho]), \;
G{\rm Tr}([\pi,\rho]^2), \;
G{\rm Tr}(\rho^2), \;
G{\rm Tr}(\rho[\rho,\rho]), \;
G{\rm Tr}([\rho,\rho]^2). \quad
\label{list}
\end{eqnarray}
Here we omit the derivatives  and also possible indices. 
There are in fact  no other couplings, and this 
is a generic result from the holographic QCD  for interactions involving
a single glueball. 
Even for multi-glueball vertices, this flavor structure is maintained.

A direct consequence of this flavor structure
is the fact that  glueballs
cannot decay to $4 \pi_0$. The decay channel to $4\pi_0$ appears in 
the $F_{\mu z}^4$ term which is in the higher DBI corrections and thus
suppressed by the large 't Hooft coupling. So, the holographic QCD
predicts 
that, among the decay products of the glueballs, $4\pi_0$ is suppressed.

It is important to note that
we work here in a ``holographic gauge'' (\ref{pirhodef})
of the D8-brane action,
in which interactions are seen in the simplest
and the most transparent way.
On the other hand, in the $A_z=0$ gauge \cite{Sakai:2004cn}
the broken chiral symmetry is manifest
because the pions fields appear in the action as $U=\exp(i \pi/ f_\pi)$.
Thus this gauge is appropriate for comparison with 
the chiral perturbation theory,
but the interactions are complicated.
A different gauge choice in the D8-brane action leads to a different
field definition of the four dimensional fields and,
of course, this does not change the physics.
Therefore, there should be some hidden structure in the QCD effective
action, 
at least in the large $N_c$ limit,
because of the higher dimensional gauge symmetry. 
For example, the vector meson dominance is its consequence 
\cite{Sakai:2005yt}.
The suppression of the $4\pi_0$ in the glueball decay
is also a consequence of this hidden structure which is not manifest
in a gauge choice other than the ``holographic gauge'' (\ref{pirhodef}).

For explicit computations, 
we concentrate on the couplings of the lightest scalar glueball, 
because of its phenomenological interest. 
But it is obvious that couplings of other glueball
excitations can be computed in the same manner. One of the
phenomenologically interesting excitations is the $2^{++}$ state. 
Fortunately, the supergravity fluctuation for this $2^{++}$ state
is simple (see for example \cite{Constable:1999gb} whose notation we
follow), in particular it consists of only the fluctuation of the metric 
components of $\mu,\nu=0,1,2,3$. So it in fact couples to the
4 dimensional part of the energy-momentum tensor of the D8-brane
Yang-Mills action. Therefore the coupling should be of the form
\begin{eqnarray}
 \int \! d^4x \; G^{\mu\nu}\eta^{\rho\sigma} 
{\rm Tr}(F_{\mu\rho}F_{\nu\sigma}),
\quad
 \int \! d^4x \; G^{\mu\nu}{\rm Tr}(\partial_\mu \pi \partial_\nu \pi),
\quad
 \int \! d^4x \; G^{\mu\nu}{\rm Tr}(\rho_\mu \rho_\nu),
\quad 
\label{glue2++}
\end{eqnarray}
at the quadratic order in $\pi$ and $\rho_\mu$. Here $F_{\mu\nu}$ is the 
field strength of the $\rho$ meson at its linear order, 
$F_{\mu\nu}\equiv \partial_\mu \rho_\nu - \partial_\nu \rho_\mu$.

Also the second lightest $0^{++}$ glueball is simple, since the
supergravity fluctuation is involved with the dilaton that has a very
simple coupling to the D8-brane Yang-Mills fields. We expect that this
kind of simplicity in the glueball couplings may give some constraint on
the decay products and decay widths, in particular the spin dependence
of the decay product, and may serve as a smoking gun for identifying 
the higher glueball states in the meson spectroscopy.\footnote{
In fact, we will find in the next subsection that there is no
mixing between the lightest $0^{++}$ glueball (or the lightest $2^{++}$)
with $q\bar{q}$ mesons, at the leading $1/\sqrt{N_c}$ order.}
In the next
subsection, we explicitly compute the interaction lagrangian of 
the lightest scalar glueball.

\subsection{Interaction of the lightest scalar glueball }

In this subsection we derive the interaction lagrangian of  glueballs
with  light $q\bar{q}$ mesons (the pions and the $\rho$
mesons). First, we need to fix the normalization of  eigenfunctions
in higher dimensions, in the dual side. Then, we substitute all the
fluctuations into the D8-brane action and perform  integration over
extra dimensions to obtain the interaction lagrangian. The interaction
lagrangian includes a possible mixing between the glueball states and
the $q\bar{q}$ mesons; however, we show that there is no mixing for the
lightest glueball.

\subsubsection{Normalization of the fluctuation fields}

Sakai and Sugimoto \cite{Sakai:2004cn,Sakai:2005yt} have determined the
normalization of the eigenfunctions $\psi_1(z)$ and $\phi_0(z)$ for the
$\rho$ meson and the pion respectively:
\begin{eqnarray}
&& \frac23 R_{\rm SS}^{9/2} U_{\rm KK}^{1/2} {\cal T}_{\rm D8} 
V_4 g_s^{-1} 
(2\pi\alpha')^2 \int \! dZ \; K^{-1/3} (\psi_1)^2 = 1, 
\label{rhonorm}\\
&&\frac32 R_{\rm SS}^{3/2} U_{\rm KK}^{7/2} 
{\cal T}_{\rm D8} V_4 g_s^{-1} \
(2\pi\alpha')^2 \int \! dZ \; K (\phi_0)^2 = 1.
\end{eqnarray}
With these, substituting (\ref{pirhodef}) into the D8-brane action
(\ref{d8ac}) (the metric is fixed with its background value), we obtain
canonically normalized kinetic terms for the $\rho$ meson and the pion,
\begin{eqnarray}
 S_0 = -{\rm Tr}\int d^4x \left\{
\frac12 (\partial_\mu \pi)^2 + \frac14 F_{\mu\nu}^2 + \frac12 \lambda_1
M_{KK}^2 \rho_\mu^2
\right\}.
\end{eqnarray}
The field strength $F$ is that of the $\rho$ meson, 
$F_{\mu\nu}\equiv \partial_\mu \rho_\nu - \partial_\nu \rho_\mu$. 

On the other hand, the normalization of the fluctuation eigenfunction
$H(Z)$ for the glueball has not been carried out in the past. The
normalization of $H(Z)$ in (\ref{glfl}) 
should be fixed in such a way that substitution of 
the expressions (\ref{glfl}) into the supergravity action (\ref{s11-7})
results in, after the integration of the extra dimensions, a canonical
kinetic term for the glueball field $G(x)$, 
\begin{eqnarray}
 S = {\rm const.} - \int\! d^4x 
\left(
\frac12 (\partial_\mu G)^2 + \frac12 M^2 G^2
\right) + {\cal O}(G^3).
\end{eqnarray}
We numerically solve the differential equation (\ref{Hd}) for $H(Z)$, to
get\footnote{Note that there are no higher derivative terms on the right
hand side of this expression. This is due to a useful gauge choice for
the gravity fluctuations
(\ref{glfl}), introduced in \cite{Constable:1999gb}.}
\begin{eqnarray}
&& \int d^7 x \sqrt{-\det G_{MN}} \left(R(G) + \frac{30}{L^2}\right)
\nonumber \\
&& \hspace{10mm}
= -0.0574\frac{R^4}{L^3}\left(H(Z=0)\right)^2
\int d^4x d\tau dx^4
\left[
\partial_\mu G \partial^\mu G
+ M^2  G^2
\right].
\end{eqnarray}
So, using the expressions for the M-theory gravity
coupling $\kappa_{11}$, and also the relations between the supergravity
parameters and the QCD parameters \cite{Sakai:2004cn,Sakai:2005yt}
\begin{eqnarray}
 R_{\rm SS}^3 = \frac12 \frac{g_{\rm YM}^2 N_c l_s^2}{M_{\rm KK}},
\quad
U_{\rm KK} = \frac29 g_{\rm YM}^2 N_c M_{\rm KK} l_s^2, \quad
g_s=\frac{1}{2\pi} \frac{g_{\rm YM}^2}{M_{\rm KK}l_s},
\label{QCDpara}
\end{eqnarray}
we obtain the normalization 
\begin{eqnarray}
 \left(H(Z=0)\right)^{-1} = 0.00978 g_{\rm YM} N_c^{3/2} M_{KK}.
\label{normH}
\end{eqnarray}

\subsubsection{Glueball interaction lagrangian}

Once the normalization of the eigenfunctions $H(Z)$, $\psi_1(Z)$ and 
$\phi_0(Z)$ are determined, substituting all the fluctuations
(+ background) into the D8-brane action (\ref{d8ac}) gives us the
glueball - $q\bar{q}$ meson couplings. We concentrate on interactions
linear in the glueball field $G$, since we are interested in the
glueball decays. 

First, because the D8-brane action is written in terms of the
type IIA string metric
and the dilaton field, we need a
dimensional reduction from the 11 dimensional fields to the 10
dimensional fields. We find 
\begin{eqnarray}
&& g_{rr} = \frac{L}{rf} 
\left(
1 + \frac{L^2}{2r^2}h_{44} + \frac{r^2 f}{L^2}h_{rr}
\right), \quad 
g_{\mu\nu} = \frac{r^3}{L^3}
\left(
\left(1 + \frac{L^2}{2r^2}h_{44}\right)\eta_{\mu\nu}
+ \frac{L^2}{r^2}h_{\mu\nu}
\right), \nonumber 
\\
&& g_{r\mu} = \frac{r}{L} h_{r\mu}, \quad 
g_{\tau\tau} =
\frac{r^3}{L^3} f
\left(
1 + \frac{L^2}{2r^2}h_{44} + \frac{L^2}{r^2 f} h_{\tau\tau}
\right), \quad 
e^{4\Phi/3} = \frac{r^2}{L^2} + h_{44}.
\end{eqnarray}
Substituting these and all the expressions for the fluctuations
(\ref{glfl}) and (\ref{pirhodef}) into the D8-brane action (\ref{d8ac}), 
we obtain the interaction action $S_{\rm int}$ before the integration
over the extra dimension $Z$: 
\begin{eqnarray}
&& \frac{-{\cal T}_{\rm D8}(2\pi\alpha')^2 V_4}{4g_s}\;{\rm Tr}\!
\int  \! d^4x dZ \left[
3 R_{\rm SS}^{3/2} U_{\rm KK}^{5/2} K
\left\{
\frac12 \left((\partial_\mu \pi)^2 \phi_0^2 + \rho_\mu^2 \psi_1^2\right)
\widetilde{H}
\left(1-\frac{\square}{M^2}\right) G 
\right.
\right.
\nonumber \\
&&
\hspace{50mm}\left.
+\left(\partial_\mu \pi\partial_\nu \pi 
\phi_0^2 + \rho_\mu \rho_\nu  \psi_1^2\right) \widetilde{H}
\frac{\partial^\mu \partial^\nu}{M^2} G 
\right\}
\nonumber \\
&&
\hspace{10mm}
+ \frac{2}{3} K^{-1/3} R_{\rm SS}^{9/2} U_{\rm KK}^{-1/2}
\left\{
-\frac12 F_{\mu\nu}^2 \psi_1^2 \widetilde{H}
\left(1+\frac{\square}{M^2}\right)G
+ 2 F_{\mu\rho}F_\nu^{\;\rho}\psi_1^2 \widetilde{H}
\frac{\partial^\mu \partial^\nu}{M^2} G 
\right\}
\nonumber \\
&&
\hspace{40mm}
\left.
- \frac{180 K}{(5K-2)^2} R_{\rm SS}^{3/2} U_{\rm KK}^{3/2} 
Z \psi_1(\partial_Z\psi_1)H
\rho_\nu F_{\mu}^{\;\nu} \frac{\partial^\mu}{M} G
\right].
\label{intac}
\end{eqnarray}
We have defined 
$\widetilde{H}(Z)\equiv ((1/4) + 3/(5K-2))H(Z)$.
In this action (\ref{intac}) we have kept
only terms quadratic in $\pi$ and $\rho_\mu$, for simplicity.
Note that the first line in the interaction action (\ref{intac})
vanishes for an on-shell glueball, $(\square - M^2)G=0$.

Terms of higher order in $\pi$ and $\rho$ in the list (\ref{list})
can also be computed in the same manner.
(Other couplings do not appear at this  order in the large
't Hooft coupling expansion.) Among these additional couplings, 
only the $G\pi\pi\rho$ coupling shown below will be relevant 
for the later computations of the decay width: 
\begin{eqnarray}
&& \frac{6i{\cal T}_{\rm D8}(2\pi\alpha')^2 V_4}{4g_s}
\int  \! d^4x dZ \left[
 R_{\rm SS}^{3/2} U_{\rm KK}^{5/2} K
{\rm Tr}
\left(\partial_\mu \pi[\rho_\nu,\pi]\right) \phi_0^2\psi_1
\widetilde{H}
\frac{\partial^\mu\partial^\nu}{M^2} G 
\right].
\label{gppr}
\end{eqnarray}
In addition to this, there is another term for $G\pi\pi\rho$ which
vanishes for the on-shell glueball.  

Finally, performing the $Z$ integration, we obtain the following
interaction lagrangian (in this expression, again we have kept 
only terms 
quadratic in $\pi$ and $\rho_\mu$):
\begin{eqnarray}
S_{\rm int} = 
&& 
-{\rm Tr}\int d^4x
\left\{
c_1\frac14 (\partial_\mu \pi)^2 \left(1-\frac{\square}{M^2}\right)G
+ c_2 
\frac14 M_{KK}^2 \rho_\mu^2
\left(1-\frac{\square}{M^2}\right)G
\right.
\nonumber \\
&&
+c_1\frac12 (\partial_\mu \pi \partial_\nu \pi)
\frac{\partial^\mu \partial^\nu}{M^2}G
+ c_2 
\frac12 M_{KK}^2 \rho_\mu \rho_\nu
\frac{\partial^\mu \partial^\nu}{M^2}G
\nonumber \\
&&
\left.
-c_3\frac18 F_{\mu\nu}^2 \left(1+\frac{\square}{M^2}\right)G
+ c_3 
\frac12 F_{\mu\rho}F_{\nu}^{\;\rho}
\frac{\partial^\mu \partial^\nu}{M^2}G
-c_4\frac32 \rho_\nu F_{\mu}^{\;\nu}\frac{\partial^\mu}{M^2}G
\right\}.
\label{sint}
\end{eqnarray}
Here $G$ is the lightest scalar glueball field with the mass $M$,  
$J^{PC}=0^{++}$, and 
$F_{\mu\nu}\equiv \partial_\mu \rho_\nu - \partial_\nu \rho_\mu$ is the
field strength of the rho meson $\rho_\mu$, and the coefficients $c_1
\sim c_4$ are defined as follows: 
\begin{eqnarray}
&& c_1 \equiv
\int dZ \frac{1}{K \pi} \widetilde{H}, \\
&&
c_2 \equiv 
\frac23 R_{\rm SS}^{9/2}U_{\rm KK}^{1/2}
{\cal T}_{\rm D8}V_4 (2\pi\alpha'^2)g_s^{-1}
\int\! dZ \;K (\partial_Z \psi_1)^2 \widetilde{H},
\\
&&
c_3 \equiv 
\frac23 R_{\rm SS}^{9/2}U_{\rm KK}^{1/2}
{\cal T}_{\rm D8}V_4 (2\pi\alpha'^2)g_s^{-1}
\int\! dZ\; K^{-1/3} (\psi_1)^2 \widetilde{H},
\\
&&
c_4 \equiv 
\frac23 R_{\rm SS}^{9/2}U_{\rm KK}^{1/2}
{\cal T}_{\rm D8}V_4 (2\pi\alpha'^2)g_s^{-1}
M_{KK}^2\int\! dZ \frac{20 K Z}{(5K-2)^2} \psi_1
(\partial_Z \psi_1) H.
\end{eqnarray}
These are evaluated numerically,
\begin{eqnarray}
 c_1=\frac{44.3}{g_{\rm YM}N_{\rm c}^{3/2}M_{\rm KK}},
\;
 c_2=\frac{5.03}{g_{\rm YM}N_{\rm c}^{3/2}M_{\rm KK}},
\nonumber \\
 c_3=\frac{49.3}{g_{\rm YM}N_{\rm c}^{3/2}M_{\rm KK}},
\;
 c_4=\frac{-0.0732 M_{\rm KK}}{g_{\rm YM}N_{\rm c}^{3/2}}.
\label{c1c4}
\end{eqnarray}
The $G\pi\pi\rho$ coupling (\ref{gppr}) is integrated to give
\begin{eqnarray}
 i c_5 \int\! d^4x\; {\rm Tr}(\partial_\mu \pi [\rho_\nu,\pi]) 
\frac{\partial^\mu \partial^\nu}{M^2}G, 
\quad {\rm with} \;
 c_5 \equiv \int \! dZ \frac{1}{K\pi} \psi_1\widetilde{H}
=  \frac{1.43\times 10^3}{g_{\rm YM}^2 N_c^{5/2}M_{\rm KK}}.
\quad
\label{c5}
\end{eqnarray}
These are the basic ingredients for computing the decay of the 
lightest glueball in section \ref{sec:3}.

\setcounter{footnote}{0}

\subsubsection{Mixing of glueball with $q\bar{q}$ mesons}

For the identification of the glueball state in the data of the 
real hadronic spectra, mixing with other states possessing the same
quantum number (for the scalar glueball of our interest, it is $0^{++}$) 
is quite essential. Generically, the mixing is expected to appear, 
because no symmetry can prohibit it. In holographic QCD,
 mixing can be computed explicitly.\footnote{See
\cite{Barbon:2004dq} for 
glueball mixing to $\eta'$ meson in a different holographic model of
QCD based on flavor D6-branes \cite{Kruczenski:2003uq}.} 
Here we show that
the lightest scalar glueball has no mixing, in the leading order 
interaction lagrangian of our concern. This means that
the mixing is largely suppressed, so the decay of the lightest scalar
glueball is dominated by a direct decay (not through the mixing of the
mass matrix). Note that for generic glueball excitations, this is not
the case, as we will see below.

First, let us give a generic argument on the mixing of a generic 
glueball and $q\bar{q}$ mesons, in the holographic QCD. 
We can show that the mixing is suppressed by
$1/\sqrt{N_c}$. 
The mixing, a linear coupling between a glueball and a $q\bar{q}$
meson,  originates in a linear coupling between supergravity fields
and Yang-Mills/scalar
fields in the D8-brane action (\ref{d8ac}). The order of the mixing can
be identified after canonically normalizing the glueball field $G$ and a
meson field $X$. We already know that the
normalization of the glueball field $G$ is given by (\ref{normH}), while
that of the $X$ meson can be characterized by that of the 
$\rho$ meson, (\ref{rhonorm}). 
Using the relations (\ref{QCDpara}), the pre-factor in
(\ref{rhonorm}) can be computed as $g_{\rm YM}^2 N_c^2/108\pi^3$. 
The D8-brane action has the same
pre-factor (since the pre-factor in (\ref{rhonorm}) is basically for
canonically normalizing the $\rho$ meson kinetic term in the D8-brane
action), so the mixing term can be written as 
\begin{eqnarray}
 S_{\rm mix} &\sim& 
\frac{g_{\rm YM}^2 N_c^2}{108\pi^3}\!
\int\! d^4x \; X G \!\int \! dZ\; \psi_1^{(X)} H 
\;\sim \;
\frac{(g_{\rm YM}^2 N_c^2)}{(g_{\rm YM}^2 N_c^2)^{1/2} 
(g_{\rm YM}N_c^{3/2})} 
\!\int\! d^4x \; X G 
\nonumber \\
&\sim& 
\frac{1}{\sqrt{N_c}}\int\! d^4x \; X G.
\end{eqnarray}
This means that the mixing is of order $1/\sqrt{N_c}$.\footnote{
This suppression can be understood more easily. 
The supergravity fields are normalized with
$1/g_s^2$ factor in front of the supergravity action, while the D8-brane 
gauge fields are normalized with $1/g_s$ factor in the tension of the
D8-brane ${\cal T}_{\rm D8}$. Therefore, if one canonically normalizes
the kinetic terms of the fluctuation fields, the above-mentioned
mixing coupling receives a $\sqrt{g_s}$ factor, and in view of the
AdS/CFT correspondence this factor is just a $1/\sqrt{N_c}$ correction.  
} 

Although it is suppressed in the large $N_c$ limit, this has a
significant effect on the decay process. 
The direct meson decay process comes from meson interactions
which are of order $g_{\rm YM}^{-1} N_c^{-1}$. So, combining this with
the mixing, the total decay amplitude through the mixing is 
$\sim g_{\rm YM}^{-1} N_c^{-3/2}$. 
On the other hand, the couplings computed 
in (\ref{c1c4}) mean that the direct decay amplitude 
of the glueball is of order $g_{\rm YM}^{-1} N_c^{-3/2}$, 
which is the same as the mixing decay amplitude.
Therefore in generic glueball decay, direct decay process is
comparable to the decay through the mixing term.  

Our interest here is primarily on the lightest glueball, and let us show
that there is no mixing for this lightest glueball. First, note that the
$\rho$ meson and the 
pion appear quadratically in the D8-brane action. This already shows
that for the glueballs originating in the dilaton and
the graviton fluctuations have no mixing with the $\rho$ meson and the
pion. (For glueballs coming from NSNS $B$-field or RR gauge fields may 
have 
mixings.) So the lightest glueball can mix only with other type of
mesons which are not in the higher dimensional Yang-Mills field; that is
the transverse scalar field $y$
on the D8-brane.
This $y$ is
not written explicitly in (\ref{d8ac}) but included in the induced
metric and the dilaton. 
The KK decomposition of the field $y$ 
produces scalar mesons with quantum number $0^{++}$. 
This $y$ is again an $N_f \times N_f$ matrix. When $N_f=3$, among
$N_f^2=9$ matrix elements, we have 
two elements with isospin zero, which mix with the glueball.
These two $q\bar{q}$ meson states can be identified with $f_0$ 
mesons (other than the glueball candidate $f_0(1500)$); 
near $f_0(1500)$, there are $f_0(1370)$ and $f_0(1710)$,
which may be identified with these two $q\bar{q}$ mesons coming from
$y$.  

Possible mixings among these three $f_0$ mesons have been studied
phenomenologically (see for example 
\cite{Amsler:1995tu,Lee:1999kv}). 
However, we show below that holographic QCD predicts there is 
no mixing at the leading order.\footnote{
The components with isospin $= 1$ are identified
as $a_0(1450)$ meson \cite{Sakai:2004cn}.}
From the induced metric in the D8-brane action, we are interested in
terms linear in the field $y$, which would lead to possible
mixing. These are 
\begin{eqnarray}
&& g_{\nu y}|_{y=0}\partial_\mu y(z,x^\mu), \quad 
 g_{\nu y}|_{y=0}\partial_z y(z,x^\mu), \quad 
 g_{z y}|_{y=0}\partial_\mu y(z,x^\mu), \quad 
 g_{z y}|_{y=0}\partial_z y(z,x^\mu), \nonumber \\
&& y [\partial_y g_{\tau\tau, rr,\mu r, \mu\nu}]_{y=0}, \quad
y [\partial_y \phi]_{y=0}, \quad
\label{linear}
\end{eqnarray}
where the last term is of course from the dilaton.
Here the bulk coordinates $y$ and $z=ZU_{\rm KK}$ are
\cite{Sakai:2004cn}  
\begin{eqnarray}
 y = \left(
U_{\rm KK}\sqrt{\frac{r^6}{R^6}-1}\right)
\cos\theta ,\quad
 z = \left(U_{\rm KK}\sqrt{\frac{r^6}{R^6}-1}
\right)\sin\theta ,\quad
\theta \equiv \frac{3}{2} \frac{U_{\rm KK}^{1/2}}{R_{\rm SS}^{3/2}}\tau.
\quad
\end{eqnarray}
It is easy to show that in fact the metrics and the dilaton
appearing in (\ref{linear})
which include the glueball fluctuations (\ref{glfl}) 
disappear at $y=0$ where the D8-brane is located,
after transforming  the metric by using the $(r,\mu,\tau)$
spacetime coordinates. So we conclude that there is no mixing of the
lightest 
glueball with mesons, at the leading order  in $1/\sqrt{N_c}$.\footnote{
Here we only consider the $U(N_f)$ singlet of the transverse scalars.
We can show that 
the other non-Abelian scalars do not mix with the lightest glueball
as in the case of the $\rho$ mesons.}
Precisely the same argument shows that the lightest 
$2^{++}$ glueball also does not participate in mixing at this order.

\setcounter{footnote}{0}
\section{Decay of the lightest scalar glueball}
\label{sec:3}

Starting from  the interaction lagrangian (\ref{sint}) and (\ref{c5}),
we can directly 
study the decay products and  their decay widths. In this
section,  we first enumerate the kinematically allowed decay processes
by analyzing the 
masses  of particles involved. This provides a 
list of decay products for the lightest glueball. We then compute the
decay widths by using the interaction lagrangians 
(\ref{sint}) and (\ref{c5}) including  explicit
numerical coefficients (\ref{c1c4}). 
Finally, we compare the widths with experimental
data for the glueball candidate $f_0(1500)$. We  find that the
prediction of  
holographic QCD  qualitatively reproduces the total
width as well as the branching ratios of the $f_0(1500)$.

\subsection{Decay products}

Let us study the kinematical constraints on the decay of the lightest
glueball in the holographic QCD. 
The lightest glueball mass $M$ is given \cite{Brower:2000rp} by 
$M = \sqrt{7.31/9} M_{KK}$, while the $\rho$ meson mass is
$m_\rho=\sqrt{\lambda_1} M_{KK} = \sqrt{0.669} M_{KK}$.
So we have a relation 
\begin{eqnarray}
 m_\rho < M < 2 m_\rho
\label{massr}
\end{eqnarray}
in the holographic QCD. This means, our lightest glueball cannot decay
to two on-shell $\rho$ mesons.\footnote{
For the most probable candidate of the
glueball $f_0(1500)$, $M=1507$MeV, so 
this mass relation is satisfied in the experimental data
($m_\rho=776$MeV).}  

In (\ref{list}), we listed all the coupling terms appearing in the
interaction lagrangian with a single glueball field $G$. From those
terms one can construct Feynman diagrams for the decay processes. 
We will work with two flavors, for definiteness. The list 
(\ref{list}) 
can be grouped into two categories, as follows:
\begin{itemize}
 \item[(i)] 
$G {\rm Tr}(\pi^2), \; G{\rm Tr}(\pi,[\pi,\rho]), \; 
G{\rm Tr}(\rho^2), \;  G\eta'\eta'$
\item[(ii)] 
$G{\rm Tr}([\pi,\rho]^2), \; G{\rm Tr}(\rho[\rho,\rho]), \;
G{\rm Tr}([\rho,\rho]^2), \; G \omega\omega$
\end{itemize}
We have written explicitly and 
separately the trace part of the $q\bar{q}$ mesons: $\eta'$ for
the pions and $\omega$ for the $\rho$ mesons. In addition to these,
there are couplings coming from the Chern-Simons term, 
\begin{itemize}
 \item[(iii)] $G{\rm Tr}(\pi\rho\rho)$,
$G\eta'{\rm Tr}(\rho\rho)$, $G\omega {\rm Tr}(\pi\rho)$,
$G \eta'\omega\omega$
\end{itemize}
for which the spacetime indices are contracted by the epsilon tensor. 
The category (i) is important for the decay processes, while 
the category (ii) and (iii) are almost irrelevant kinematically, 
since with the
couplings (ii) and (iii) the final decay product includes more than 
5 pions (or 4 pions and one $\eta'$). To understand this, note that
$\rho$ meson can decay to two pions, and $\omega$ decays to three
pions\footnote{We don't consider coupling to photons in this paper.}. 
When the number of the pions are
large, typical momentum for the final pion state is small.
Pion couplings are accompanied by derivatives, then the
amplitude is expected to be suppressed. By this reason, we restrict our
analysis to the cases where the final decay product is induced by the
couplings (i).

Since the mass of the glueball is not larger than twice the $\rho$ meson
mass, the final decay products should have less than two $\rho$ mesons. 
All possible decay chains obtained by these couplings 
are categorized by the decay products:
\begin{itemize}
 \item[(a)] $ G \rightarrow \pi \pi $ \quad (figure 1)
 \item[(b)] $ G \rightarrow \rho \pi \pi$, \quad
$ G \rightarrow \rho \rho\rightarrow \rho\pi\pi$ \quad (figure 2)
 \item[(c)] $ G \rightarrow \rho \pi \pi\rightarrow \pi\pi\pi\pi$,
\quad
 $ G \rightarrow \rho\rho \rightarrow \pi \pi\pi\pi$ \quad (figure 3)
 \item[(d)] $ G \rightarrow \eta'\eta'$ \quad (figure 1)
\end{itemize}
If we think of $G$ as the $f_0$(1500), then this list 
is consistent with what is known in the particle data book
\cite{Yao:2006px}.  The branching ratio given in \cite{Yao:2006px} is
35$\%$ for (a), 49$\%$ for $G\rightarrow 4\pi$ (corresponding to
(b)+(c)\footnote{ 
In \cite{Yao:2006px}, (b) is not explicitly written, 
but we interpret that (b) is included in the $G\rightarrow 4\pi$ decay
in \cite{Yao:2006px}
because the on-shell $\rho$ meson in (b)
would decay to $2 \pi$.}),
and 7 $\%$ for (d).\footnote{
We are working in the case of two flavors, so we don't distinguish 
$\eta$ and $\eta'$, and ignore $K$. } 
So we can reproduce the main decay channels of the  $f_0(1500)$.
In the next subsection, we compute the decay widths for each of these
decay branches.

\subsection{Decay widths}

Let us evaluate the decay widths for these groups. Groups  (a) and (d)
are two-body decays so that the decay widths can be computed
analytically. For the remaining groups, (b) 
and (c), integrations over  final momenta are  complicated that we
computed the decay widths numerically.\footnote{
We worked in the supergravity convention, but to compute the 
decay widths it is convenient to go to the metric convention
$(+1,-1,-1,-1)$. The new action $S = S_0 + S_{\rm int}$ quadratic in
the ${q\bar{q}}$ meson fields is
\begin{eqnarray}
 S_0 = &&{\rm Tr}\int\! d^4x \left\{
\frac12 (\partial_\mu \pi)(\partial^\mu \pi) 
- \frac14 F_{\mu\nu}F^{\mu\nu} + \frac12 \lambda_1
M_{KK}^2 \rho_\mu \rho^\mu
\right\},
\\
S_{\rm int} = 
&& 
{\rm Tr}\int\! d^4x
\left\{
c_1\frac14 (\partial_\mu \pi)(\partial^\mu \pi)
 \left(1+\frac{\square}{M^2}\right)G
+ c_2 
\frac14 M_{KK}^2 \rho_\mu \rho^\mu
\left(1+\frac{\square}{M^2}\right)G
\right.
\nonumber \\
&&
-c_1\frac12 (\partial_\mu \pi \partial_\nu \pi)
\frac{\partial^\mu \partial^\nu}{M^2}G
- c_2 
\frac12 M_{KK}^2 \rho_\mu \rho_\nu
\frac{\partial^\mu \partial^\nu}{M^2}G
\nonumber \\
&&
\left.
+c_3\frac18 F_{\mu\nu}F^{\mu\nu} \left(1-\frac{\square}{M^2}\right)G
+ c_3 
\frac12 F_{\mu\rho}F^{\nu\rho}
\frac{\partial^\mu \partial_\nu}{M^2}G
+c_4\frac32 \rho_\nu F^{\mu\nu}\frac{\partial_\mu}{M^2}G
\right\}.
\end{eqnarray}
}

\subsubsection{$G \rightarrow \pi \pi$, $G\rightarrow \eta\eta$}

Two-body decays are the simple to analyze, for which we have for the
decay width,  
\begin{eqnarray}
 \Gamma = \frac{|{\bf p}|}{8\pi M^2} |{\cal M}|^2,
\end{eqnarray}
where ${\cal M}$ is the amplitude of the graph responsible for the
decay, and ${\bf p}$ is the final momentum of one of the identical
particles in the decay product. 

\begin{figure}[t]
\begin{center}
 \begin{minipage}{15cm}
\begin{center}
\includegraphics[width=4cm]{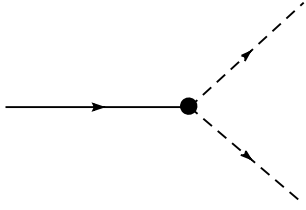}
\caption{A glueball $G$ decaying to two pions $\pi$. }
\label{fig-gpp}
\end{center}
 \end{minipage}
\end{center}
\end{figure}

In the rest frame of the glueball, the first line in the interaction
lagrangian (\ref{sint}) vanishes. For the $2\pi$ decay, the relevant
coupling in that frame is
\begin{eqnarray}
 \frac12 c_1 \partial_0 \pi^a \partial_0 \pi^a G.
\end{eqnarray}
For definiteness we consider a specific 
adjoint index for the pion $\pi^a$ ($a=1,2,3$). We have two pions
as a final state, $\pi^{(1)}$ and $\pi^{(2)}$, then
\begin{eqnarray}
 {\cal M} = \frac{1}{2} c_1 i p_0^{\pi^{(1)}}
i p_0^{\pi^{(2)}} \times 2,
\end{eqnarray}
where the last factor 2 is for the symmetry of exchanging 
the two final identical particles.The kinematics shows that
$p_0^{\pi^{(1)}} =  p_0^{\pi^{(2)}} = |{\bf p}| = M/2$
because the pions are massless, 
so we obtain 
\begin{eqnarray}
 {\cal M} = -c_1 M^2/4.
\end{eqnarray}
The decay width summed over $a=1,2,3$ is
\begin{eqnarray}
 \Gamma_{G\rightarrow \pi\pi} 
= \frac{ {|c_1|^2}M^3}{256 \pi}\times 3 \times \frac12 .
\end{eqnarray}
The last factor $1/2$ is necessary because the final state has
two identical particles. 

For evaluating the numerical value of the decay width, 
we need $\kappa \equiv \lambda N_c / 108\pi^3 = 7.45 \times 10^{-3}$
which was used in \cite{Sakai:2004cn} to fit the pion decay constant, 
and $N_c=3$. Using these as inputs, we finally obtain the decay width
\begin{eqnarray}
 \frac{\Gamma_{G\rightarrow\pi\pi}}{M} = 0.040.
\end{eqnarray}
This is to be compared with the experimental data in \cite{Yao:2006px},
\begin{eqnarray}
\frac{\Gamma^{(\rm ex)}_{G\rightarrow \pi\pi}}{M}  
= \frac{109}{1507} \times 34.9\% = 0.0252,
\end{eqnarray}
with which we find a qualitatively good agreement.

Another two-body decay channel is for $G\rightarrow\eta\eta$. The
$\eta'$ mass evaluated in the holographic QCD in
\cite{Sakai:2004cn}\footnote{The expression is 
$m_{\eta'} = \frac{1}{3\sqrt{6}\pi}\sqrt{\frac{N_f}{N_c}}
(g_{\rm YM}^2 N_c)M_{\rm KK}\sim 17.4M_{\rm KK}$.
} 
is found to be too large, $2m_{\eta'}> M$, so in the holographic QCD
this decay channel cannot be described. However, if we adopt, as a
trial, the $\eta'$ mass as a free parameter in this holographic QCD, 
then we obtain the decay width
\begin{eqnarray}
 \frac{\Gamma_{G\rightarrow\eta\eta}}{M} = 
\frac{\Gamma_{G\rightarrow\pi\pi}}{M} \times \frac13 \times
\sqrt{1-\frac{4m_\eta^2}{M^2}}.
\end{eqnarray}
The factor $1/3$ is to suppress the effect of the three kinds of the
pions, and the last factor is necessary to replace $|{\bf p}|$ of the
pion with that of the $\eta$ meson. If we substitute the real
observed ratio 
$m_\eta/M_{f_0(1500)}=547.5/1507$,  we obtain
\begin{eqnarray}
  \frac{\Gamma_{G\rightarrow\eta\eta}}{M} = 0.0090.
\end{eqnarray}
We compare this with the experimental data, 
\begin{eqnarray}
\frac{\Gamma^{(\rm ex)}_{G\rightarrow \eta\eta}
+ \Gamma^{(\rm ex)}_{G\rightarrow \eta\eta'}}{M}  
= \frac{109}{1507} \times 7.0\% = 0.00506,
\end{eqnarray}
again this is qualitatively in agreement with our result.

\subsubsection{$G \rightarrow \rho \pi \pi$}

\begin{figure}[t]
\begin{center}
 \begin{minipage}{14cm}
\begin{center}
\begin{tabular}{ccc}
\includegraphics[width=4cm]{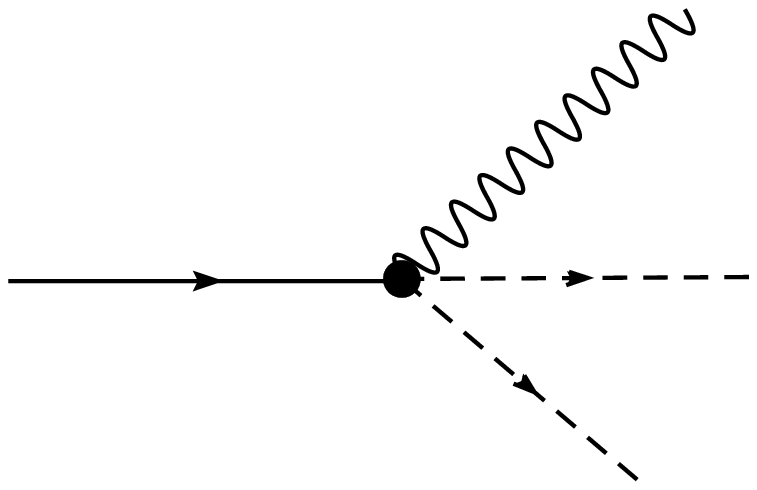}
&\ \ &
\includegraphics[width=4cm]{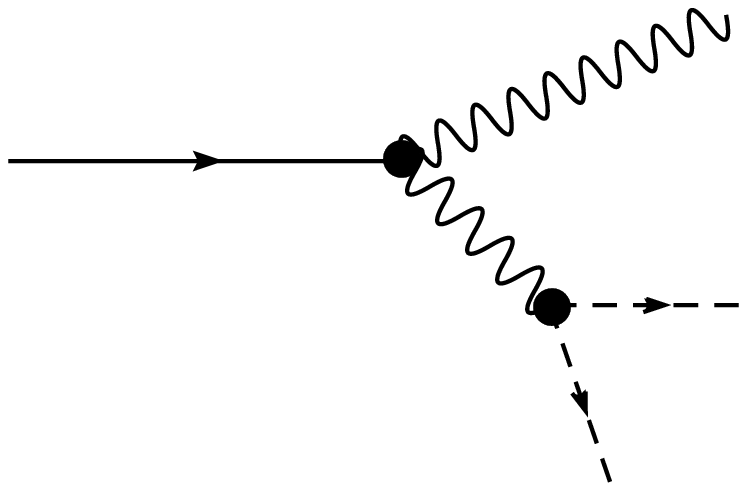}\\
\end{tabular}
\put(0,-5){$\pi^1$}
\put(-25,-30){$\pi^3$}
\put(-60,-8){$\rho^2$}
\put(0,40){$\rho^2$}
\put(-100,5){$G$}
\put(-250,-15){$G$}
\put(-150,5){$\pi^1$}
\put(-175,-20){$\pi^3$}
\put(-150,40){$\rho^2$}
\caption{A glueball $G$ decaying to two pions $\pi$ and a single
 $\rho$. There are two graphs, the decay with a single vertex (Left)
and the decay with two vertices (Right).
}
\label{fig-grpp}
\end{center}
 \end{minipage}
\end{center}
\end{figure}

First we describe the decay $G \rightarrow \rho \pi \pi$
which uses the single vertex (\ref{c5}), see figure \ref{fig-grpp}
(Left). 
The $SU(2)$ generators are normalized as $\sigma^a /\sqrt{2}$, so the
interaction (\ref{c5}) is written explicitly as
\begin{eqnarray}
-\sqrt{2} c_5 \epsilon_{abc}
\partial_\mu \pi^a \rho_\nu^b \pi^c
\frac{\partial^\mu \partial^\nu}{M^2} G
\;=\;
\sqrt{2} c_5 \epsilon_{abc}
\partial_0 \pi^a \rho_0^b \pi^c G
\end{eqnarray}
where we have used a relation in the rest frame of the glueball $G$, 
\begin{eqnarray}
 \frac{\partial_\mu \partial_\nu}{M^2} G
= \frac{\delta^\mu_0 \delta^\nu_0 (ip_G^0)^2}{M^2} G
= -\delta^\mu_0 \delta^\nu_0 G.
\end{eqnarray}
Labeling the decay products as 
$\pi^1({\bf p}_{(1)}) + \rho^2_\nu({\bf p}_{(\rho)}) 
+ \pi^3({\bf p}_{(2)})$, the amplitude is
\begin{eqnarray}
 {\cal M}= \sqrt{2}c_5 \bigl(i p_{(1)0} - ip_{(2)0} \bigr)
\epsilon_0 
\end{eqnarray}
where 
$\epsilon_0 = |{\bf p_{(\rho)}}|/m_\rho$ is the zeroth component of the
$\rho$ meson polarization vector, 
\begin{eqnarray}
 \epsilon^\mu = \left(
\frac{|{\bf p_{(\rho)}}|}{m_\rho},
\frac{-p_{(\rho)0}}{|{\bf p_{(\rho)}}|m_\rho} {\bf p_{(\rho)}}
\right).
\label{polar}
\end{eqnarray}
Other polarization vectors have vanishing $\epsilon_0$.

Next, we evaluate the amplitude for the process 
$G\rightarrow \rho\rho\rightarrow \rho\pi \pi$, see
figure \ref{fig-grpp} (Right).
Let us list the Feynman rules: The 
$\pi^a \rho_\nu^b \pi^c G$ vertex is the same as before,
$\sqrt{2}c_5 i p_0^{(\pi^a)} \delta^\nu_0 \epsilon_{abc}$, and 
the $\pi^d \rho_\mu^e \pi^f$ vertex was obtained in \cite{Sakai:2004cn}
as 
\begin{eqnarray}
 \sqrt{2}c_6 i p_\mu^{(\pi^d)} \epsilon_{def}
\quad
{\rm where}
\quad 
 c_6 \equiv \int_{-\infty}^{\infty}
dZ \frac{1}{\pi K} \psi_1
= \frac{24.0}{N_c g_{\rm YM}}.
\end{eqnarray}
The $\rho_\mu^a\rho_\nu^a G$ vertex (no sum over $a$) is
\begin{eqnarray}
&& \frac12 c_2 M_{KK}^2
\delta^\mu_0\delta^\nu_0
- \frac12 c_3 (p_{(\nu)}^\sigma p_{(\mu)\sigma}\eta^{\mu\nu} 
- p_{(\mu)}^\nu p_{(\nu)}^\mu)
\nonumber \\
&&
+\frac12 c_3 \left(
p_{(\mu)0} p_{(\nu)0}\eta^{\mu\nu}
-p_{(\mu)}^\nu p_{(\nu)0}\delta^\mu_0
-p_{(\nu)}^\mu p_{(\mu)0}\delta^\nu_0
+p_{(\mu)}^\sigma p_{(\nu)\sigma}\delta^\nu_0\delta^\mu_0
\right)
\nonumber \\
&&
-\frac34 c_4 \frac{1}{M}
\left(
p_{(\nu)0}\eta^{\mu\nu} - p_{(\nu)}^\mu \delta^\nu_0
+p_{(\mu)0}\eta^{\mu\nu} - p_{(\mu)}^\nu \delta^\mu_0
\right), 
\end{eqnarray}
and the $\rho$ meson propagator is
\begin{eqnarray}
 \frac{1}{p_{(\rho)}^2 - m_\rho^2 + i m_\rho \Gamma_\rho}
\left(
\delta^\mu_\nu - \frac{p^\mu_{(\rho)} p_{(\rho)\nu}}{m_\rho^2}
\right)\delta^{be}.
\end{eqnarray}
Here we need the $\rho$ meson decay width, 
$\Gamma_\rho/m_\rho=c_6^2/24\pi$, which can be evaluated as
$\Gamma_\rho/m_\rho = 0.307$ in the holographic QCD
\cite{Sakai:2004cn,Sakai:2005yt}.  
This is close to the experimental
value $\Gamma_\rho/m_\rho = 149.4/775.5=0.1927$ in \cite{Yao:2006px}.
As for the polarization vectors, in addition to (\ref{polar}), we have
two more vectors. One is
\begin{eqnarray}
 \epsilon^\mu= \left(
0,\frac{p^{(\rho)}_2}{\sqrt{(p^{(\rho)}_1)^2 + (p^{(\rho)}_2)^2}},
\frac{-p^{(\rho)}_1}{\sqrt{(p^{(\rho)}_1)^2 + (p^{(\rho)}_2)^2}},0
\right)
\label{polar2}
\end{eqnarray}
and the other gives the same decay width
with that of the polarization (\ref{polar2}). 
Using all of these Feynman rules, we can compute the decay amplitude 
for the process $G\rightarrow \rho\rho\rightarrow \rho\pi \pi$. 

The total expression for the decay width is lengthy, and we here provide 
only the numerical results after substituting the necessary inputs used
also in the evaluation of the $\Gamma_{G\rightarrow \pi \pi}$. 
We obtain
\begin{eqnarray}
\frac{\Gamma_{G\rightarrow\rho\pi\pi}}{M} =  3 \times 
\left(
1.1 \times 10^{-8} + 2 \times (2.1\times 10^{-7})
\right)\sim 1.3 \times 10^{-6}.
\label{resgrpp}
\end{eqnarray}
The  factor $3$ is for  the sum over possible flavor indices. 
(The decay width we computed is 
for the decay $G \rightarrow \pi^1 \rho^2 \pi^3$,
and there are two other possible combinations for the indices.)
Terms in the parentheses are for different polarizations of the
final $\rho$ meson.

Our result (\ref{resgrpp}) is very small. 
The smallness mainly comes from the fact that 
the integration region of the momentum is so small 
because $m_\rho$ is very close to $M$ in our holographic computation. 
In reality, the mass of $f_0(1500)$ is much larger than the mass of the
$\rho$ meson. 
So, as a trial, in our computation of the decay width,
let us modify the input glueball mass $M$ such that $M/m_\rho$ 
coincides with the experimental value in \cite{Yao:2006px}. 
Then, we obtain
\begin{eqnarray}
\frac{\Gamma_{G\rightarrow\rho\pi\pi}}{M} =  0.096.
\label{fitrpp}
\end{eqnarray}
Since the decay product $\rho\pi\pi$ seems to be included in
$G\rightarrow 4\pi$ in \cite{Yao:2006px}, we compare our result with the
experimental value after adding the decay width of $G\rightarrow 4\pi$
which we compute next.

\setcounter{footnote}{0}

\subsubsection{$G\rightarrow 4 \pi$}

\begin{figure}[t]
\begin{center}
 \begin{minipage}{13cm}
\begin{center}
\begin{tabular}{ccc}
\includegraphics[width=4cm]{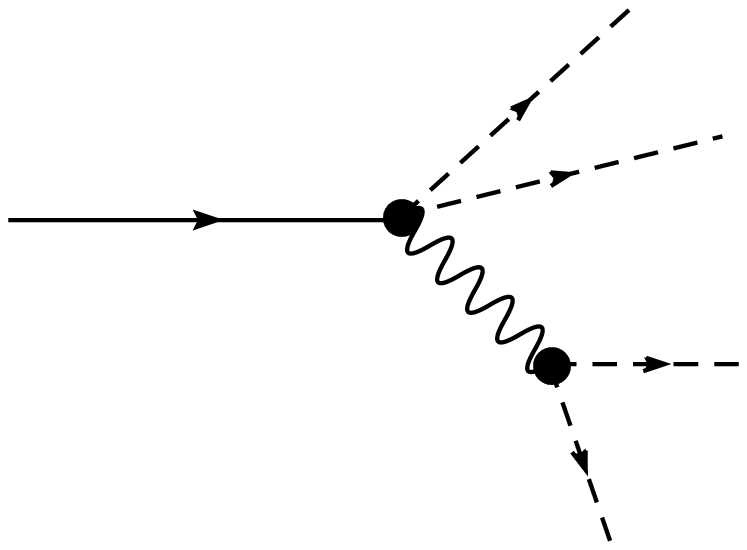}
&\ \ &
\includegraphics[width=4cm]{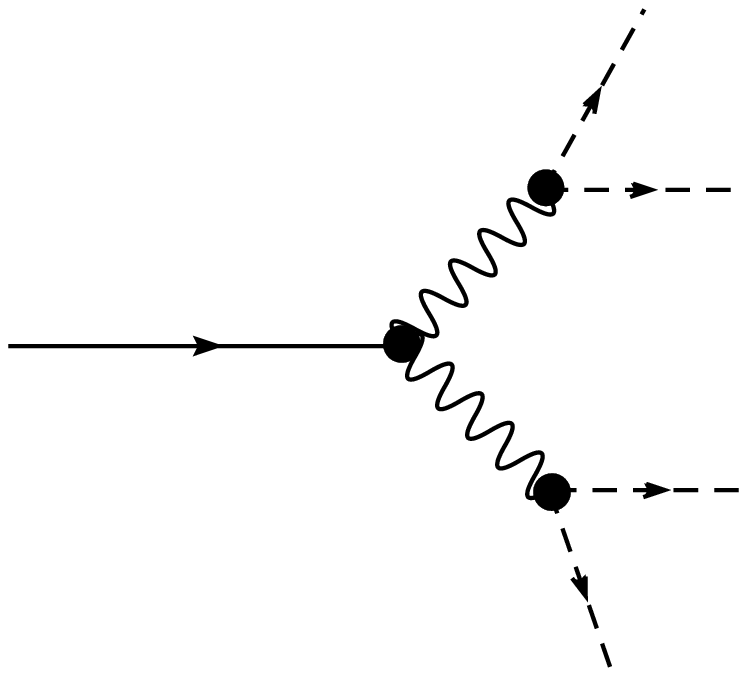}\\
\end{tabular}
\put(-5,-20){$\pi^d$}
\put(-25,-40){$\pi^f$}
\put(-65,-15){$\rho^e$}
\put(-60,20){$\rho^b$}
\put(-5,30){$\pi^c$}
\put(-20,50){$\pi^a$}
\put(-100,10){$G$}
\put(-250,-10){$G$}
\put(-147,15){$\pi^c$}
\put(-145,-20){$\pi^d$}
\put(-163,40){$\pi^a$}
\put(-165,-45){$\pi^f$}
\put(-205,-15){$\rho^e$}
\caption{A glueball $G$ decaying to four pions $\pi$. There are two
 graphs, the decay with two vertices (Left) 
and the decay with three vertices (Right).
}
\label{fig-g4p}
\end{center}
 \end{minipage}
\end{center}
\end{figure}

The computation of this amplitude is done in the same manner, and we do 
not write it explicitly, except for some important points. 
First, it is easy to find out that the decay product is only in the
combination $G \rightarrow 2\pi^i 2\pi^j \; (i\neq j)$, 
So one can specifically choose $i=1,j=2$ for the computation.
(This determines the index for the $\rho$ meson as $\rho^3$.)
Since the amplitude is proportional to 
\begin{eqnarray}
\delta^{be}\epsilon_{abc}\epsilon_{def} = \delta_{ad} \delta_{cf}
-\delta_{af}\delta_{cd},
\end{eqnarray}
all possible ways to assign the adjoint index for each of the
final pions are 

\vspace{5mm}
\begin{tabular}{c|cccccccccccccccc}
$\pi^1({\bf p}_{(1)})$ & a & d& a &d&a&f&a&f&c&f&c&f&c&d&c&d\\
$\pi^1({\bf p}_{(2)})$ & d & a& d &a&f&a&f&a&f&c&f&c&d&c&d&c\\
$\pi^2({\bf p}_{(3)})$ & c & c& f &f&c&c&d&d&a&a&d&d&a&a&f&f\\
$\pi^2({\bf p}_{(4)})$ & f & f& c &c&d&d&c&c&d&d&a&a&f&f&a&a\\
\hline
sign & + & + &+&+&$-$&$-$&$-$&$-$& + & + &+&+&$-$&$-$&$-$&$-$
\end{tabular}
\vspace{5mm}

\noindent
We have to sum the amplitude with these substitutions, with the sign
indicated in the table.

The main difficulty in the computation resides in the evaluation of the
integration of the final momenta, $\int d^3{\bf p}_{(1)}d^3{\bf p}_{(2)}
d^3{\bf p}_{(3)}d^3{\bf p}_{(4)}$. In the integrand of the decay width, 
there is
a four dimensional delta function coming from the total energy momentum
conservation. 
The integration $\int d^3{\bf p}_{(4)}$ trivially 
eliminates three delta functions
of the momentum conservation. Furthermore, using spatial
rotation symmetry, we can orient one of the remaining momenta as
$\vec{{\bf p}}_{(3)} = (p,0,0)$,
then the remaining single delta function for the energy conservation 
can be eliminated by the integration 
$\int d^3{\bf p}_{(3)} = 4 \pi \int p^2 dp$. 
Specifically, the delta function 
is expressed as
\begin{eqnarray}
&& \delta\left(
p + \sqrt{p^2 + 2p (p_{(1)x}+p_{(2)x}) 
+ |{\bf p}_{(1)} + {\bf p}_{(2)}|^2}
+|{\bf p}_{(1)}| + |{\bf p}_{(2)}|-M
\right)
\nonumber \\
&&
= 
\frac{2(M-|{\bf p}_{(1)}|-|{\bf p}_{(2)}|)(p_{(1)x}+p_{(2)x}) + 
|{\bf p}_{(1)}+{\bf p}_{(2)}|^2 + (M-|{\bf p}_{(1)}|-|{\bf p}_{(1)}|)^2}
{2(M-|{\bf p}_{(1)}|-|{\bf p}_{(2)}|+p_{(1)x}+p_{(2)x})^2}
\nonumber \\
&&
\hspace{5mm} 
\times \delta
\left(
p-
\frac{
 (M-|{\bf p}_{(1)}|-|{\bf p}_{(2)}|)^2-
|{\bf p}_{(1)}+{\bf p}_{(2)}|^2 }
{2(M-|{\bf p}_{(1)}|-|{\bf p}_{(2)}|+p_{(1)x}+p_{(2)x})}
\right).
\end{eqnarray}
So the integration over $p$ results in a constraint for the remaining
momenta ${\bf p}_{(1)}$ and ${\bf p}_{(2)}$.  This constraint
corresponds to a restriction on  the integration 
region, $|{\bf p}_{(1)}|+|{\bf p}_{(2)}|+
|{\bf p}_{(1)}+{\bf p}_{(2)}| \leq M$.

Numerical integration for the remaining momenta gives the decay
width,\footnote{This  value includes a factor $3$ accounting for
different  combinations of the species of the final decay product.
}
\begin{eqnarray}
 \frac{\Gamma_{G\rightarrow 4\pi}}{M} \sim 2.2\times 10^{-5}.
\end{eqnarray}
If we adjust the glueball mass to the experimental value (while fixing
the $\rho$ meson mass by $m_\rho = \sqrt{\lambda_1}M_{\rm KK}$)
as was done before, we obtain 
\begin{eqnarray}
 \frac{\Gamma_{G\rightarrow 4\pi}}{M} \sim 0.0087.
\label{fit4p}
\end{eqnarray}

Let us compare our results (\ref{fitrpp}) and (\ref{fit4p}) with the
experimental values. Adding these two, we obtain
\begin{eqnarray}
 \frac{\Gamma_{G\rightarrow 4\pi}
+\Gamma_{G\rightarrow \rho\pi\pi}
}{M} \sim 0.105,
\label{decaypi4}
\end{eqnarray}
while the experimental data \cite{Yao:2006px} shows
\begin{eqnarray}
\frac{\Gamma_{G\rightarrow 4\pi}^{\rm (ex)}}{M} 
=\frac{109}{1507}\times 49.5\% = 0.0358.
\end{eqnarray}
One can see that the order of this decay width is reproduced in the
holographic QCD.\footnote{
If we include the mass of the pions in some way, 
the decay width (\ref{decaypi4}) to the four pions 
is expected to become significantly smaller and close to the
experimental data.}

\section{Concluding remarks}
\label{sec:4}

We have presented here  the first attempt in computing decays of 
glueballs to $q\bar{q}$ mesons using  holographic QCD. 
We have adopted a string-theoretic set-up, (which is of the so-called
``top-down'' type), the 
Sakai-Sugimoto model. The glueball sector lives in supergravity 
fluctuations in the Witten's background of non-BPS black 4-branes, and
the mesons live on the probe D8-branes. The coupling between the two
sectors is encoded in the D8-brane action, and KK decomposition and
integration over extra dimensions gives the desired couplings in four
spacetime dimensions. 

Explicit couplings between the lightest glueball and the $q\bar{q}$
mesons are given, and the associated  decay products/widths
are calculated. We find that our results are consistent with 
the experimental data of the decay for the $f_0(1500)$ which is thought
to be the best candidate of a glueball in the hadronic spectrum.

The most important 
merit of the holographic QCD is that one  can go beyond the chiral
perturbation theory; one can compute coefficients which cannot be fixed
solely by the chiral symmetry. At low energy the chiral perturbation
works well, but at the energy scale of the glueball mass the derivative
expansion in the chiral perturbation becomes unreliable. Furthermore, 
glueballs are flavor-blind, so it is quite difficult to constrain
possible interactions from the chiral symmetry. The holographic
description obtained in the holographic QCD is,
in principle, equivalent to QCD, though in the large $N_c$ and large
't Hooft coupling limit.\footnote{Precisely speaking, the limit to
QCD includes $M_{\rm KK}\to\infty$ (there should be a double-scaling
limit with a simultaneous scaling of the 't Hooft coupling), 
after incorporating infinite number of $1/N_c$ corrections. The
background receives large stringy corrections and becomes essentially a
purely stringy background.} 
We therefore expect that the holographic approach
should provide interesting information on strong coupling physics of
QCD. In 
fact, we have discussed generic features of the glueball interactions
predicted in holographic QCD (see section \ref{sec:gene}). For example, 
we have argued that, among the
decay products of glueballs, $4 \pi_0$ should be suppressed. 

One of the reasons why the $f_0(1500)$ is expected to be 
a glueball state is that the $f_0(1500)$ does not decay to $2\gamma$. 
In the holographic QCD, we can compute relevant photon coupling in
the same manner, and find that $G\gamma\gamma$ coupling is vanishing at
the leading order (see \cite{Sakai:2004cn,Sakai:2005yt} for the way to
introduce the electromagnetic field as an external background of the
massless fields on the D8-branes). Since we have shown that there is no
mixing with $q\bar{q}$ mesons at the leading order, our result of the
suppressed photon coupling reproduces the experimental data.

In this paper, we have explicitly computed  for the decay of the
lightest glueball, which 
is of the most phenomenological interest. There are also many other
interesting directions,  e.g., generalizing our results using
 various approaches in holographic QCD. Here are some examples:
\begin{itemize}
 \item 
Multi-glueball couplings. Self-couplings of the glueballs 
can be computed in the supergravity sector. Emission of mesons from
a propagating glueball can be described by the D8-brane action
similarly. 
For highly spinning glueballs whose holographic dual are closed fundamental
strings in the confining supergravity background, their decay into two
glueballs was briefly described in \cite{Cot}.
\item Universally narrow width of glueballs.
If one can show in the holographic QCD 
that the total decay width of any glueball state is narrow, 
that would provide support for  this widely-held belief.  In this paper
       we have shown the narrowness 
only for the lightest glueball. Explicit calculation of the widths 
is possible for other
glueball excitations, as they are available in
       \cite{Brower:2000rp}. The $2^{++}$  glueball coupling 
has been described in (\ref{glue2++}) for example. 
\item Glueballs with other $J^{PC}$ quantum numbers.
Glueball states
       originating in the RR fields in the supergravity may possess
       interesting structure in the meson couplings. The $0^{-+}$
       glueball is described by a RR 1-form $C_\tau$ whose fluctuation
       is completely decoupled from the others, and it appears in the
       Chern-Simons coupling in the D8-brane action. The $1^{+-}$ 
       glueballs reside in the NS-NS 2-form field, and it should have 
 a large
       mixing with the meson fields. This is a consequence of the gauge
       invariance in the supergravity, requiring the gauge-invariant
       combination $B_{\rm NSNS} + F$ in the D8-brane action.
\item Thermal/dense QCD. One can modify the supergravity background, or
       introduce a background for the D8-brane fields, to describe the
       finite temperature/baryon density\footnote{There are many recent
       works on introducing finite temperature or finite baryon density
       (or chemical potential) in holographic models of
       QCD. Specifically, for the Sakai-Sugimoto model, see
       \cite{Aharony:2006da}.}, that surely will modify the couplings
       which we have computed. Glueball couplings should be sensitive to
       the deconfinement tempearture, and, near the transition
       temperature, they should become singular in some sense. This is
       of phenomenological interest in view of the onset of the LHC. 
\item Computation of the glueball couplings in other models of
       holographic QCD. For a single-flavor case, flavor D6-branes 
enable one to introduce easily the quark mass \cite{Kruczenski:2003uq}, 
which might shed some light on how our results may be modified by the
       pion masses. 
To apply our strategy to so-called bottom-up
       phenomenological approach in holographic QCD may reveal how
       universal the glueball couplings obtained in our paper are.
\end{itemize}
All of these would be interesting to investigate.

\acknowledgments 
K.~H.~would like to thank T.~Sakai and S.~Sugimoto for valuable 
comments, and like to thank Brown university high energy theory group
for kind invitation. S.~T.~ is grateful to D.~Jido and M.~Murata for
useful discussions. 
K.~H.~and S.~T.~thank the Yukawa Institute for Theoretical Physics at
Kyoto University which supported the YITP workshop YITP-W-07-05 on
``String Theory and Quantum Field Theory'' during which valuable 
discussions on
the present paper were made.
K.~H.~and S.~T.~are partly supported by
the Japan Ministry of Education, Culture, Sports, Science and
Technology. 


\end{document}